\documentclass[aps,pra,amsmath,amssymb,preprint,superscriptaddress,longbibliography, nofootinbib]{revtex4-1}
\usepackage{amsmath}
\usepackage{amsthm}
\usepackage{subcaption}
\usepackage{amsfonts}
\usepackage{amssymb}
\usepackage[pdftex]{xcolor,graphicx}
\usepackage{color}
\usepackage[utf8]{inputenc}
\usepackage{hyperref} %para ligar las referencias
\usepackage[english]{babel}
\usepackage{natbib}
\usepackage{array}
\usepackage[T1]{fontenc}
\usepackage{float}
\usepackage{placeins}
\DeclareUnicodeCharacter{00A8}{\textasciitilde}

\begin{document}
	
%--------------------------------------------------------------------------------
	\title{Squeezed displaced entangled states in the quantum Rabi model}
	\author{F. H. Maldonado-Villamizar}
	\email[e-mail: ]{fmaldonado@inaoep.mx}
	
	\affiliation{CONACYT-Instituto Nacional de Astrof\'isica, \'Optica y Electr\'onica, Calle Luis Enrique Erro No.
		1, Sta. Ma. Tonantzintla, Pue. CP 72840, M\'exico}

	\author{C. Huerta Alderete}

	\affiliation{Instituto Nacional de Astrof\'isica, \'Optica y Electr\'onica, Calle Luis Enrique Erro No.
		1, Sta. Ma. Tonantzintla, Pue. CP 72840, M\'exico}
	
	\author{B. M. Rodr\'iguez-Lara}
	\affiliation{Tecnologico de Monterrey, Escuela de Ingenieria y Ciencias, Ave. Eugenio Garza
	Sada 2501, Monterrey, N.L., 64849, M\'exico}
	\affiliation{Instituto Nacional de Astrof\'isica, \'Optica y Electr\'onica, Calle Luis Enrique Erro No.
	1, Sta. Ma. Tonantzintla, Pue. CP 72840, M\'exico}	
	
	\begin{abstract}
	The quantum Rabi model accepts analytical solutions in the so-called degenerate qubit and relativistic regimes with discrete and continuous spectrum, in that order. 
	We show that solutions are the superposition of even and odd displaced number states, in the former, and infinitely squeezed coherent states, in the latter, of the boson field correlated to the internal states of the qubit.
 	We propose a single parameter model that interpolates between these discrete and continuous spectrum regimes to study the spectral statistics for first and second neighbor differences before the so-called spectral collapse.
 	We find two central first neighbor differences that interweave and fluctuate keeping a constant second neighbor separation.
	\end{abstract}
	
	%\pacs{05.45.Mt, 42.50.Ct, 42.50.Mn, 73.43.Nq}
	
	\maketitle
%--------------------------------------------------------------------------------
%%%%%%%%%%%%%%%%%%%%%%%%%%  body  %%%%%%%%%%%%%%%%%%%%%%%%%%
\section{Introduction}
\noindent
The quantum Rabi model (QRM) is the lowest dimensionality Hamiltonian describing light-matter interaction,
\begin{equation}\label{eq:QuRabiModel}
\hat{H}=\omega \hat{a}^{\dagger}\hat{a}+\frac{\omega_0}{2} \hat{\sigma}_z+g(\hat{a}^{\dagger} +\hat{a})\sigma_x.
\end{equation}
It describes a boson field, with frequency $\omega$ and represented by the annihilation (creation) operator $\hat{a}$ ($\hat{a}^{\dagger}$), interacting with a qubit, with frequency $\omega_{0}$ and represented by Pauli matrices $\sigma_{j}$ with $j=x,y,z$. 
Trapped ions \cite{Lv2018p021027} and superconducting circuits \cite{Mezzacapo2014p7482} provide highly controllable experimental platforms for the quantum simulation of the model in the different interaction regimes defined by the coupling strength to field frequency ratio \cite{Casanova2010p263603,Yoshihara2016p44,Kockum2919p19}. 

Through its history, the QRM has motivated the development of computational tools for both spectral and dynamic calculations \cite{Wolf2012p053817,Maciejewski2014p16}.
The model conserves parity and is solvable \cite{Braak2011p100401,Moroz2013p319}. 
It is possible to diagonalize it in the qubit basis \cite{Moroz2014p252,Moroz2016p50004}.
In the so-called adiabatic approximation, it is possible to estimate the spectrum and its eigenstates \cite{Irish2007p173601,Shen2016p044002,Xie2019p013809}. 
Generalizations that account for asymmetry between so-called rotating and counter-rotating terms as well as driving showed the existence of degeneracies in the spectrum \cite{Li2015p454005}.
Extension for more than one qubit \cite{Chilingaryan2013p335301,Wang2014p54001,Rodrguez2014p135306,Zhang2015p013814,Chilingaryan2015p245501} or field \cite{Travenec2012p043805,HuertaAlderete2016p414001,Cong2017p063803,Cong2019p013815} have been constructed.
The latter was used for the simulation of para-particles in trapped-ion setups \cite{HuertaAlderete2017p013820,HuertaAlderete2017p043835,HuertaAlderete2018p11572}.

The QRM accepts analytic solutions in the so-called degenerate qubit, $\omega_{0}=0$, and relativistic, $\omega = 0$, regimes \cite{Pedernales2015p15472}.
After diagonalization in the qubit basis, the first is reduced to two decoupled harmonic oscillators with discrete spectrum and the second to a Dirac equation in (1+1)D with continuous spectrum.
In the following, we introduce a single-parameter QRM that interpolates between these two regimes, diagonalize it in the qubit basis, and write its eigenvalue problem in the Bargmann representation.
Our model is engineered to show spectral collapse \cite{Ng1999p119,Hwang2015p180404,Duan2016p464002,Penna2017p045301}.
It is well known that the eigenstates in the degenerate qubit regime are the superposition of even and odd displaced number states correlated to the ground and excited state of the qubit, in that order. 
For the sake of completeness, we revisit this result.
Then, we calculate the expected but unreported eigenstates in the relativistic regime, which are the unbalanced superposition of position states that can be written as infinitely squeezed coherent states correlated to the internal states of the qubit.
Afterwards, we numerically explore the statistics of the first and second neighbors spectral differences before the transition to the continuous spectrum regime.
Their histograms show the interweaving of two central separations for nearest neighbors that keep a constant second neighbor spectral separation.
These results correlate with Braak conjectures regarding the spectrum of the QRM in the Bargmann representation \cite{Braak2011p100401}.
Finally, we use Husimi Q-function to visually explore the displacement and squeezing of the ground state of our model as the control parameter takes us close to the continuous spectrum regime. 

%\subsection{applications of rabi spectrum}

\section{Model}
We propose a single-parameter QRM,
\begin{equation}\label{eq:002}
H(\delta)=2\omega(1-\delta)\hat{a}^{\dagger}a+\delta \omega_{0}\hat{\sigma}_z+g(\hat{a}^{\dagger}+\hat{a})\hat{\sigma}_x,
\end{equation}
that interpolates between the so-called degenerate qubit, $\omega_{0} = 0$, and relativistic, $\omega = 0 $, regimes for the extremal values of the control parameter $\delta\in\left[0,1\right]$.
We recover the QRM for $\delta=1/2$. 
A trapped ion quantum simulation \cite{Pedernales2015p15472}, 
$ \hat{H}= (\delta_{b}-\delta_{r}) \hat{a}^{\dagger}a / 2 -  (\delta_{b} + \delta_{r}) \frac{\hat{\sigma}_z}{4}  + \Omega (\hat{a}^{\dagger}+\hat{a}) \hat{\sigma}_{x}, $
provides the desired level of control through blue (red) driving sidebands, $\delta_{b}=\omega_{b,r}-\omega_{0} - \nu$ ($\delta_{r}=\omega_{r}-\omega_{0} +\nu$) where the parameter $\omega_{b}$ ($\omega_{r}$) is the frequency of the driving laser detuned to the blue (red) of the ion transition frequency $\omega_{0}$, and the center of mass motion frequency is given by $\nu$.
The effective coupling strength $\Omega$ is related to the Lamb-Dicke parameter of the trap and the amplitude of the driving field.

\begin{figure}[htbp!]
\includegraphics{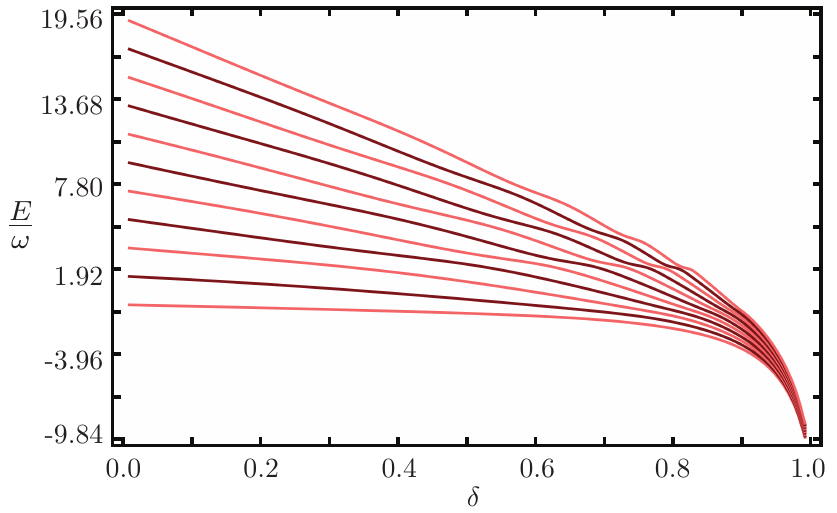}
	\caption{First eleven pseudo-energies of $\hat{H}_{+}(\delta)$ for parameters $\delta \in [0, 1)$ and $\{ \omega_{0}, g \} = \{ 1, 0.99 \} \omega$.} \label{fig:Fig1}
\end{figure}

Strictly speaking, our model is solvable \cite{Braak2011p100401,Guan2018p315204}.
We focus on the regimes with analytic closed form solution and favour the Fulton-Gouterman procedure to diagonalize it in the qubit basis $\{\vert +\rangle,\vert -\rangle\}$ \cite{Moroz2013p319},
\begin{equation}\label{eq:004}
\hat{H}_{FG}(\delta)=\hat{H}_{+}(\delta)\vert +\rangle\langle +\vert +\hat{H}_{-}(\delta)\vert-\rangle\langle -\vert,
\end{equation}
where the dynamics in the boson sector,
\begin{eqnarray}\label{eq:004a}
\hat{H}_{\pm}(\delta)=2\omega(1-\delta)\hat{a}^{\dagger} a+g(\hat{a}^{\dagger}+\hat{a})\mp\delta \omega_0\hat{\Pi},
\end{eqnarray} 
include the boson parity operator, $\hat{\Pi}=e^{i\pi \hat{a}^{\dagger} a}$. 
In the following sections, we discuss the closed form analytic solutions in the extremal regimes and
provide a numerical study for the transition from discrete to continuous spectrum, Fig. \ref{fig:Fig1}.

In this frame, we use Bargmann representation \cite{Braak2011p100401}, $a^{\dagger}\to z$ and $a\to d/dz$ with $z\in\mathbb{C}$, to write the eigenvalue problem $\hat{H}_{\pm} \psi_{\pm} = E_{\pm} \psi_{\pm}$ in differential form, 
\begin{align}\label{eq:004b}
&( z+\tilde{g})\frac{d}{dz}\phi_{1}(z)-x \phi_{1}(z)=\tilde{\omega_{0}}\phi_{2}(z) \\
&( z-\tilde{g})\frac{d}{dz}\phi_{2}(z)+(2\tilde{g}^{2}-x-2\tilde{g}z)\phi_{2}(z)=\tilde{\omega_{0}}\phi_{1}(z),
\end{align}
where we obviate the subindices and use the auxiliary functions $\phi_{1,2}=e^{\tilde{g} z-g^{2}}\psi(z)$ in terms of the eigenfunction $\psi(z)$.
We define scaled coupling and frequency parameters $\tilde{g}=g/\left[2(1-\delta)\omega \right]$ and $\tilde{\omega_{0}}=\delta\omega_{0}/\left[2(1-\delta)\omega \right]$. 
For the sake of space, we define an auxiliary scaled displaced energy $x=\tilde{g}^{2}+\tilde{E}$ in terms of the scaled eigenenergy $\tilde{E}= E / \left[2(1-\delta)\omega \right] $. 
In this representation, the parity operator acts as  $\hat{\Pi}\psi(z)=\psi(-z)$. 
A series solution of the auxiliary functions, $\phi_{2}(x;z)=\sum_{j=0}^{\infty}K_{j}(x)(z+\tilde{g})^{j}$ which implies $\phi_{1}(x;z)=\sum_{j=0}^{\infty}K_{j}(x)\tilde{\omega_{0}}(z+\tilde{g})^{j} / {(j-x)}$, 
allows one  to construct an analytic function $G(x;z)=\phi_{2}(-z)-\phi_{1}(z)$.
As the auxiliary functions $\phi_{1,2}(\pm z)$ are series expansion of the same function outside the poles, the analytic function $G(x;z)$ must vanish for true solutions for any value of $z$.
In particular, when $z=0$ we have
\begin{align}\label{eq:004c}
G(x;0)&=\sum_{j=0}^{\infty}K_{j}(x)\left(1-\frac{\tilde{\omega_{0}}}{j-x}\right)\tilde{g}^{j},
\end{align}
where the $j$-th coefficient is given by a three-term recurrence relation,
\begin{align}
jK_{j}(x) =K_{j-1}(x)f_{j-1}(x)-K_{j-2}(x)
\end{align}
with initial terms $K_{0}(x)=1$ and $K_{1}(x)=f_{0}(x)$.
The auxiliary function is given in the following,
\begin{align}
f_{j}(x)&=\frac{1}{2\tilde{g}} \left(j-x+4\tilde{g}^{2}-\frac{\tilde{\omega_{0}}^{2}}{j-x}\right).
\end{align}
The solution to the eigenvalue problem reduces to calculate the roots of $G(x;0)$ as Braak demonstrated \cite{Braak2011p100401}. 
This cannot be addressed in a direct way but Braak conjectured that the distance between poles is given by the scaled displaced energy $x$, and that zero, one or two roots can exist between consecutive poles without successive occurrences of two roots.
In the following, we will argue that these conjectures are related to the first and second neighbor energy separation.

\section{Degenerate qubit regime}

It is well known that in the degenerate qubit regime, $\delta \rightarrow 0$, the boson sector Hamiltonian reduces to a driven harmonic oscillator,
\begin{eqnarray}
\hat{H}_{I\pm} \equiv \lim_{\delta \rightarrow 0} \hat{H}_{\pm}(\delta) = 2\omega \hat{a}^{\dagger} a+ g(\hat{a}^{\dagger}+\hat{a}),
\end{eqnarray}
that is diagonalized by a displacement,
\begin{eqnarray}
\hat{D}(\alpha)\hat{H}_{FG}(0)\hat{D}^{\dagger}(\alpha)=2\omega \hat{a}^{\dagger} a- \frac{g^{2}}{2\omega}.
\end{eqnarray}
The eigenstates are displaced number states, $\vert n, \alpha \rangle \equiv \hat{D}(\alpha) \vert n \rangle$, with equally spaced spectrum $E_n^{\alpha}=2\omega n-\frac{g^2}{2\omega}$. We used the displacement operator $\hat{D}(\alpha) = e^{\alpha \hat{a}^{\dagger}-\alpha^{*}\hat{a}}$ with parameter $\alpha=-g/2\omega$.  
The eigenstates in the laboratory frame,
\begin{eqnarray}\label{eq:007}
\vert \psi_{n} \rangle = \frac{1}{2} \left( \vert \phi_{+} \rangle \vert + \rangle + \vert \phi_{-} \rangle \vert -\rangle \right),
\end{eqnarray}
have the familiar form of a maximally entangled boson-qubit state between the unnormalized even (odd) displaced number states, 
\begin{align}\label{eq:008}
\vert \phi_{\pm}\rangle =  \left[ \hat{D}^{\dagger}(\alpha)\mp(-1)^{n}\hat{D}(\alpha) \right]  \vert n \rangle,
\end{align}
and the excited (ground) qubit state.
This is not a Schrödinger cat state \cite{Schrodinger1935p807,Brune1992p5193,Monroe1996p1131}. 
The latter needs semi-classical states in the boson sector.
We have highly non-classical boson states that belong to different parity sectors.
Quantum entangled states with parity properties are useful to detect weak forces \cite{Gilchrist2004pS828}.	

\begin{figure}[htbp!]
		\includegraphics[width=\columnwidth]{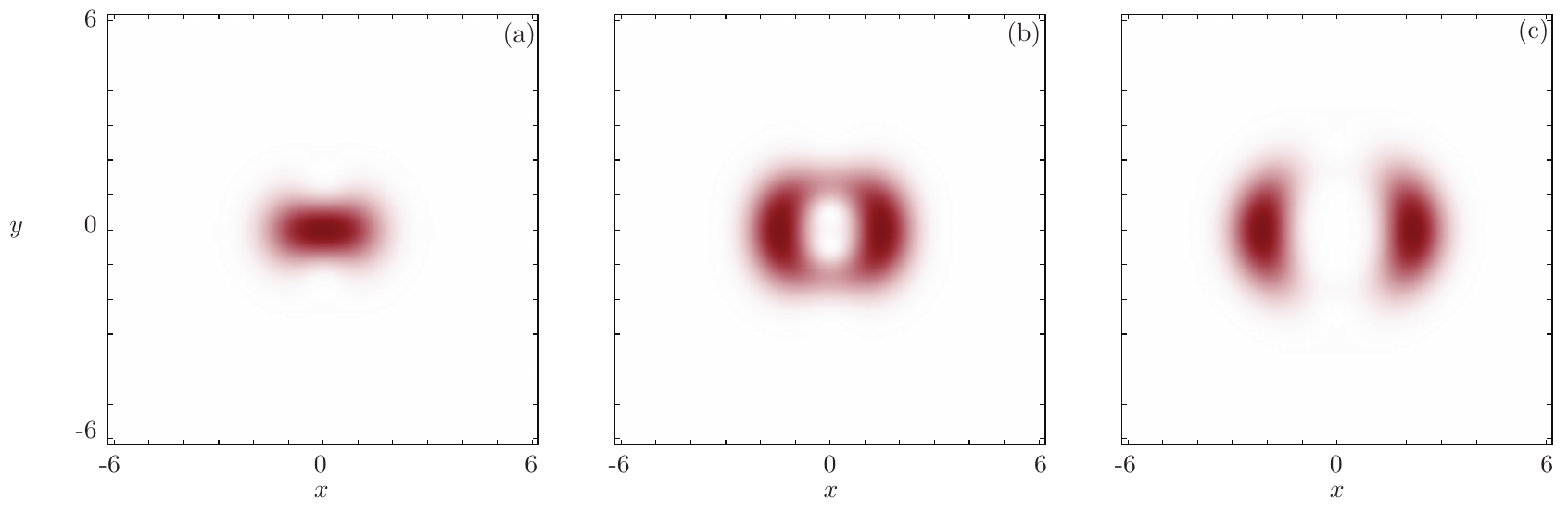}
	\caption{ Husimi Q-function for the (a) ground state, $n=0$, (b) fifth, $n=5$, and (c) tenth, $n=10$, excited states  for parameters $\delta=0$ and $\{ \omega_{0}, g \} = \{ 1, 0.99 \} \omega$.}\label{fig:Fig2}
\end{figure}

For the sake of visualization, we calculate Husimi Q-function \cite{Husmi1940p1940264} for the reduced boson sector states, 
\begin{widetext}
\begin{eqnarray}
\text{Q} (\beta) = \frac{e^{-(\vert\beta\vert^{2}+\vert\alpha\vert^{2})}}{4\pi} \sum_{r=1,-1}\left\vert \sum_{k=0}^{\infty} \frac{\beta^{*k}}{\sqrt{k!}} P(n,k,\alpha)\left[ (-1)^{\vert n - k \vert }-r(-1)^{n} \right] \right\vert^{2}
\end{eqnarray}			

where the probability of finding $k$ excitations,
\begin{eqnarray}\label{husimi}
P(n,k,\alpha) = 
 (-1)^{\vert n - k \vert \frac{(1+\text{sgn}(n-k))}{2}} \frac{ \alpha^{\vert n - k \vert} }{ \vert n - k \vert !} \sqrt{\frac{ \mathrm{max}(n,k)! }{\mathrm{min}(n,k)!}}\text{U}(-\mathrm{min}(n,k);\vert n - k \vert+1;\alpha^{2})
\end{eqnarray}
\end{widetext}
is given in terms of the Tricomi function $\mathrm{U}(n_1;n_2;x)$ \cite{VillanuevaVergara2015p22836} and the function $\mathrm{max}(n,k)$ ($\mathrm{min}(n,k)$) yields the larger (smaller) value between $n$ and $k$.
Figure \ref{fig:Fig2} shows the Husimi Q-function for the ground state, $n=0$, fifth, $n=5$ and tenth, $n=10$, excited states. 
The mean expectation value for the boson oscillator quadratures in optical phase space, $\hat{x} = \left( \hat{a}^{\dagger} + \hat{a} \right) / 2$ and $\hat{y} =  i \left( \hat{a}^{\dagger} - \hat{a} \right) / 2$, is zero and their squared standard deviations are not:
\begin{equation}
\begin{aligned}
\langle \hat{x} \rangle &= 0 , \qquad \Delta x =  \alpha^2 + \frac{n}{2}+\frac{1}{4} , \\
\langle \hat{y} \rangle &= 0,  \qquad \Delta y =\frac{n}{2}+\frac{1}{4}. 
\end{aligned}
\end{equation}
One dispersion is always larger, $\Delta x > \Delta y$,  as the displacement parameter  $\alpha$ is never zero.

\section{Relativistic regime}

Here the QRM reduces to a $(1+1)$D  Dirac equation \cite{Gerritsma2010p463,Noh203p040102,RodriguezLara2014p015803,GutierrezJauregui2018p104001} and yields an effective boson sector Hamiltonian,
\begin{equation}\label{eq1}
\hat{H}_{ R \pm} =\lim\limits_{\delta\to 1}\hat{H}(\delta)=g(\hat{a}^{\dagger}+\hat{a})\mp \omega_0\hat{\Pi},
\end{equation}
that can be mapped into a simpler form in optical phase space,
\begin{equation}\label{eq:0014}
\hat{H}_{R \pm} = 2 g \hat{x} \mp \omega_0 \hat{\Pi}.
\end{equation}
It is not bounded from below, has a continuous spectrum, and the solution is Dirac delta normalizable.
Its eigenstates in the laboratory frame, 
\begin{equation}
\begin{aligned}
\vert \psi_{0} \rangle &=  \frac{1}{\sqrt{2}} \vert 0 \rangle \left( \vert + \rangle + \vert - \rangle \right), \\
\vert \psi_{x} \rangle &= \vert \varphi_{+} \rangle \vert + \rangle + \vert \varphi_{-} \rangle \vert -\rangle,
\end{aligned}
\end{equation}
include boson sector eigenstates,
\begin{equation}
\vert \varphi_{\pm} \rangle\hspace*{-0.1cm}=\hspace*{-0.1cm}\lim\limits_{\mu \rightarrow 1} (\gamma_{+}\mp\gamma_{-}) e^{\frac{1}{4} \xi }e^{(1-\mu)\frac{\zeta^{2}}{2}} \hat{S}(-r)\hspace*{-0.1cm}\left[ D(\zeta) \mp D^{\dagger}(\zeta)\right] \vert  0 \rangle, 
\end{equation}
that are the superposition of squeezed coherent states \cite{Satyanarayana1985p400,Nieto1993p2843,Bishop1994p4488} with symmetric displacement with respect to the origin, infinite squeezing, but different probability amplitude, $\xi=\ln{(1-\mu^2)}$, $\zeta=2xe^{\frac{\xi}2{}}$ and $\mu=\tanh{(r)}$. 
Squeezed displaced states  are  interesting from a fundamental point of view \cite{Satyanarayana1985p400,Bishop1994p4488,Nieto1993p2843}.
They are a resource for high precision measurements \cite{Huang2015p17894,Kienzler2015p53,Lo2015p336} as they increase the signal-to-noise ratio \cite{Korobko2017p143601}.

In order to obtain this result, we start from the effective boson sector Hamiltonian and construct the eigenvalues for its square, $E^{2}(x)= 4 g^2 x^2+\omega_{0}^{2}$.
The square root of these eigenvalues provide a guide to construct the eigenstates using the invariant subspace spanned by the position states $\{ \vert \pm x \rangle \}$,
\begin{equation}
\begin{aligned}
\vert \Psi_{0}^{\pm} \rangle_{FG} &= \vert x=0 \rangle, \\
\vert \Psi_{x}^{\pm} \rangle_{FG} &= \frac{\omega_{0} \vert x \rangle \mp \left[ E(x) - 2 g x \right] \vert - x \rangle }{\sqrt{2E(x)} \sqrt{ E(x)-2 gx}}                                                                             , \quad x > 0.
\end{aligned}
\end{equation}
We use the map of Fock states to optical phase space, $\langle x \vert  n \rangle = \pi^{-1/4} e^{-x^{2}/2} / (\sqrt{2^{n} n!} )  \mathrm{H}_{n}(x)$, in terms of Hermite polynomials \cite{Lebedev1972p}, and translate them into operator form,  $e^{-\hat{a}^{\dagger 2} + 2 x \hat{a}^{\dagger}} \vert 0 \rangle = \sum_{n=0}^{\infty} \mathrm{H}_{n}(x) / \sqrt{n!} ~\vert n \rangle$,
\begin{equation}\label{eq:0038}
\begin{aligned}
\vert \Psi_{x}^{\pm} \rangle_{FG} &=  \left(\gamma_{+} e^{-\frac{1}{2}\hat{a}^{\dagger 2} +2 x\hat{a}^{\dagger}} \pm \gamma_{-} e^{-\frac{1}{2}\hat{a}^{\dagger 2} -2 x\hat{a}^{\dagger}} \right) \vert  0 \rangle, \\
\left\{ \gamma_{+},\gamma_{-} \right\} & = \frac{e^{-x^{2}}}{\sqrt{2 \pi E(x) \left[E(x)-2 gx \right]  }} \left\{ \omega_{0},- E(x) + 2 g x \right\}.
\end{aligned}
\end{equation}
Considering the canonical pair, $\hat{q} = \sqrt{ 2\hbar/(m \nu)}~ \hat{x}$ and  $\hat{p} = \sqrt{  2\hbar m \nu } ~\hat{y}$, where the parameters of the physical oscillator are the mass of the ion $m$ and the natural frequency $\nu$ of the ion trap, provides all the scaling factors.
The nature of the solution does not allow us to provide reliable numerics for the exact parameter value $\delta = 1$.

We can write Husimi Q-function for the reduced density matrix of the boson field,
	\begin{equation}
	\begin{aligned}
	\text{Q}(\beta) =\frac{1}{\pi^{5/4}}  \lim_{\mu \rightarrow 1}e^{\frac{1}{2} \xi -\zeta^{2}\mu}e^{-\vert\beta\vert^{2}-\vert\zeta\vert^{2}} \left[\left(\gamma_{+}-\gamma_{-}\right)^{2}\left\vert \sum_{n,m}G\left(-\mu,2m+1,2n+1,\frac{3}{2}\right)\right\vert^{2} \right.\\+\left.\left(\gamma_{+}+\gamma_{-}\right)^{2}\left\vert \sum_{n,m}G\left(-\mu,2n,2m,\frac{1}{2}\right)\right\vert^{2}   \right],
	\end{aligned}
	\end{equation}
where we construct a closed form for the weight factor \cite{Satyanarayana1985p400},

\begin{equation}
\begin{aligned}
G(\mu,m,n,\frac{1}{2}+s) &=\frac{\beta^{*n}\zeta^{*m}}{\sqrt{m!}\sqrt{n!}}\\&\times i^{m+n+s(m+1)}\left(\frac{2}{\mu}\right)^{s-\frac{m+n}{2}}\frac{\sqrt{\left(\mu^{2}-1\right)^{s+\frac{1}{2}}}}{\Gamma\left(\frac{n+3}{2}-s\right)\Gamma\left(\frac{m+3}{2}-s\right)}\\&
~~ _2F_1\left(-\frac{m+s}{2},-\frac{n+s}{2};\frac{1}{2}+s;\frac{1-\mu^{2}}{\mu^{2}}\right),
\end{aligned}
\end{equation}

in terms of Gauss hypergeometric function $\,_2F_1\left(l,k;r;x\right)$ \cite{Lebedev1972p}.
The only eigenstate that can be calculated exactly is the bosonic vacuum, $x=0$ and $\delta = 1$.

\section{Numerical analysis}

Calculating the spectrum is an important task. 
Once the spectrum is known, the intricacies of the system are revealed; for example, it is straightforward to calculate its time evolution, find its phase-space configurations \cite{Rodriguez2018p043805}, or even model its interaction with an environment \citealp{GonzalezGutierrez2017p0153301}.
The QRM is one of a few systems that can be solved \cite{Braak2011p100401}.
Here, we use numerical diagonalization to observe the changes in the spectrum of our single parameter model as it gets closer to the continuous spectrum regime. 
In our simulations we use a Fock basis of dimension $30\,000$ for the bosons on resonance with the qubit, $\omega = \omega_{0}$, an ultra-strong coupling parameter, $g=0.99 ~\omega$, and work only in the bosonic sector related to the excited qubit state in the Foulton-Gouterman reference frame.
This is equivalent to working in the positive parity subspace of the model in the laboratory reference frame.
Similar results are obtained for the negative parity subspace and different coupling parameter values.

\begin{figure}[htbp!]
\includegraphics[width=\textwidth]{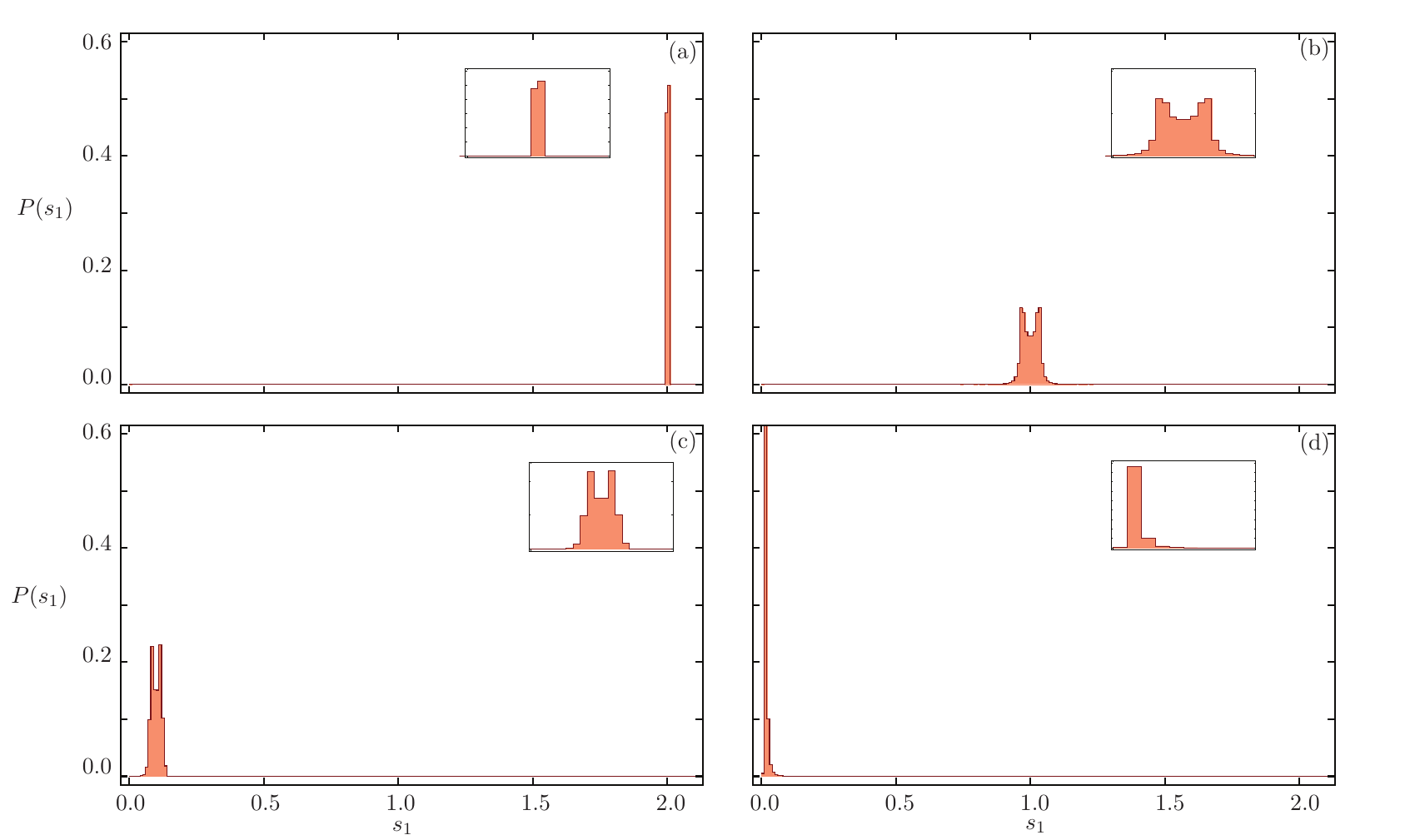}
	\caption{Histogram for the probability distribution $P(s_{1})$ for control parameter values  $\delta=$ (a) 0.0, (b) 0.5, (c) 0.95, and (d) 0.9999; all other parameters follow Fig.~\ref{fig:Fig1}. }\label{fig:Fig3}
\end{figure}

We focus on the statistic of the normalized energy separation between $k$-th nearest neighbors, $s_{k}= (E_{n+k}-E_{n})/ \hbar \omega$.
Figure \ref{fig:Fig3} shows the histogram for the first nearest neighbor probability distribution, $P(s_{1})$. 
In the degenerate qubit regime, the energy levels are equidistant and a peak at the value $ s_{1} = 2 $ appears, Fig. \ref{fig:Fig3}(a).
As the control parameter increases, the histogram displaces towards the origin, two well-defined peaks appear, and the height decreases.
For the standard QRM on resonance, $\delta = 1/2$ and $\omega = \omega_{0}$, the histogram is centered at the value $ s_{1} = 1$, Fig. \ref{fig:Fig3}(b). 
As the control parameter increases, the double peak displaces towards the origin.
At some critical value of the control parameter, the peak separation starts decreasing and the height increases, Fig. \ref{fig:Fig3}(c). 
As we get closer to the relativistic regime, where the spectrum is continuous, we expect a single peak at the value $s_{1} = 0$, Fig. \ref{fig:Fig3}(d).

We can correlate this result to Braak conjecture regarding the distance between poles and the distribution of zero, one, and two roots between consecutive poles.
First, let us discuss the separation between poles.
It is controlled by an integer auxiliary scaled displaced energy $x_{k} = \left[ 2 (1-\delta) E_{k} + g^2 \right] / \left[ 2 (1-\delta) \right] = k$.
In the degenerate case, the energy separation between  nearest spectral neighbors is two as the auxiliary scaled energy is  $x_{k} = (2 \omega E_{k} + g^2)/ ( 4 \omega^2 ) = k$.
As the control parameter $\delta$ increases, the separation between consecutive nearest spectral neighbors diminishes until, in the relativistic regime, it becomes zero.
This correlates with the transition from discrete to continuous spectrum and the displacement of the histogram towards the origin, Fig. \ref{fig:Fig3}. 
Now, the second conjecture, that there exists zero, one, or two roots between consecutive poles, can be related to the double peaked histogram.
Sections between consecutive poles that possess two roots will define short nearest spectral neighbor separation. 
Most probably, the long separation is related to the fact that consecutive two roots occurrences are not feasible.

\begin{figure}[htbp!]
	\includegraphics[width=\textwidth]{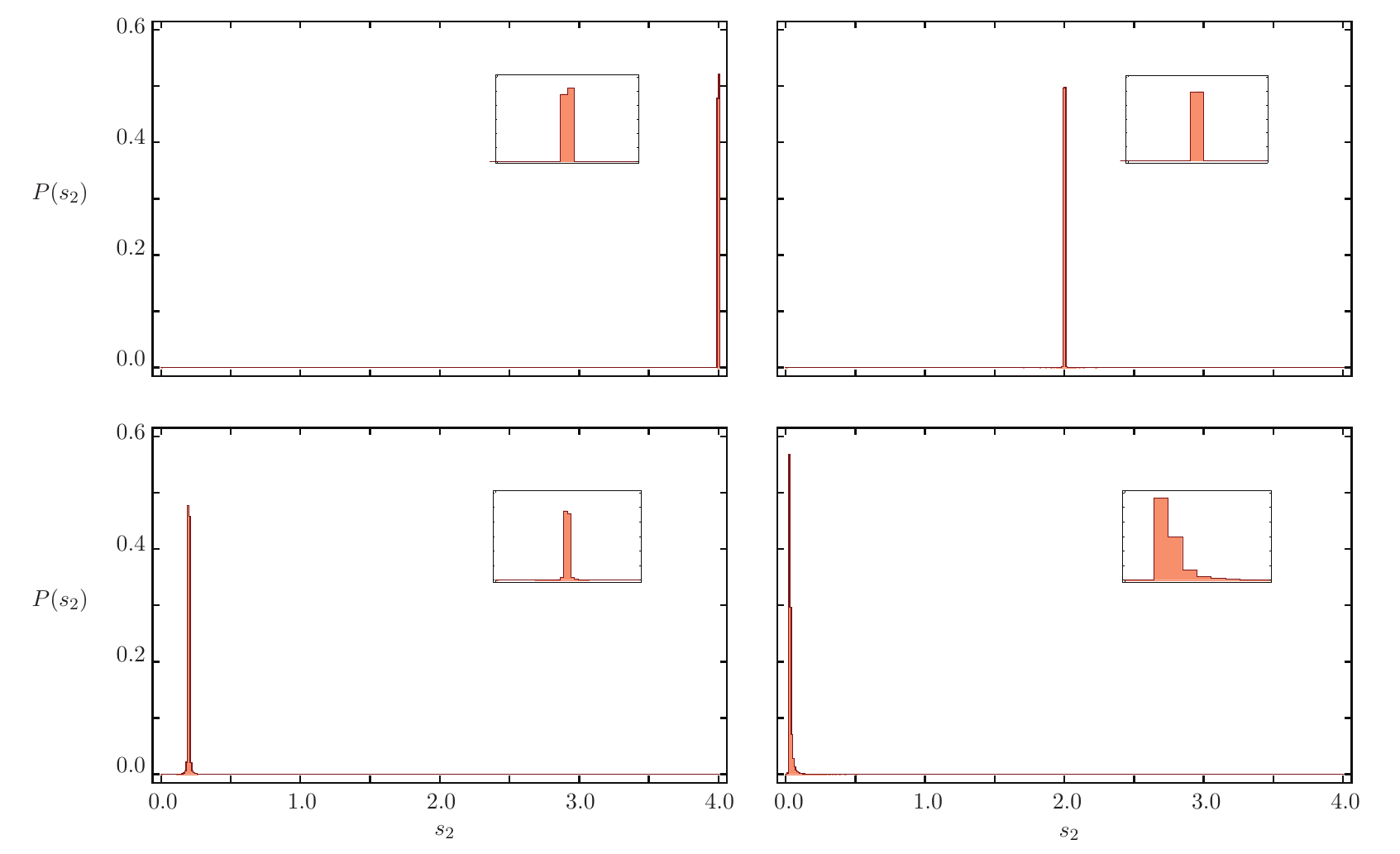}
	\caption{Histogram for the probability distribution $P(s_{2})$ for identical parameters to those in Fig. \ref{fig:Fig3}.}\label{fig:Fig4}
\end{figure}

An interesting feature arises for second nearest neighbors, $s_{2}$, where the spectral separation seems constant. 
It starts in the degenerate qubit regime as a single peak at the value $s_{2} = 4$, Fig. \ref{fig:Fig4}(a), that displaces towards the origin at almost constant height, Fig. \ref{fig:Fig4}(b) to Fig. \ref{fig:Fig4}(c), that should become a single peak at $s_{2}=0$ as we get closer to the relativistic regime, Fig. \ref{fig:Fig4}(d). 
It is straightforward to relate the displacement of the histogram with the separation between poles but we cannot provide a viable conjecture for the constant separation at the moment. 
As we said before, first neighbor separation points to the existence of two central separations, one short and one large, that interweave and fluctuate, keeping a constant second neighbor separation.

\begin{figure}[htpb!]
	\includegraphics[width=\textwidth]{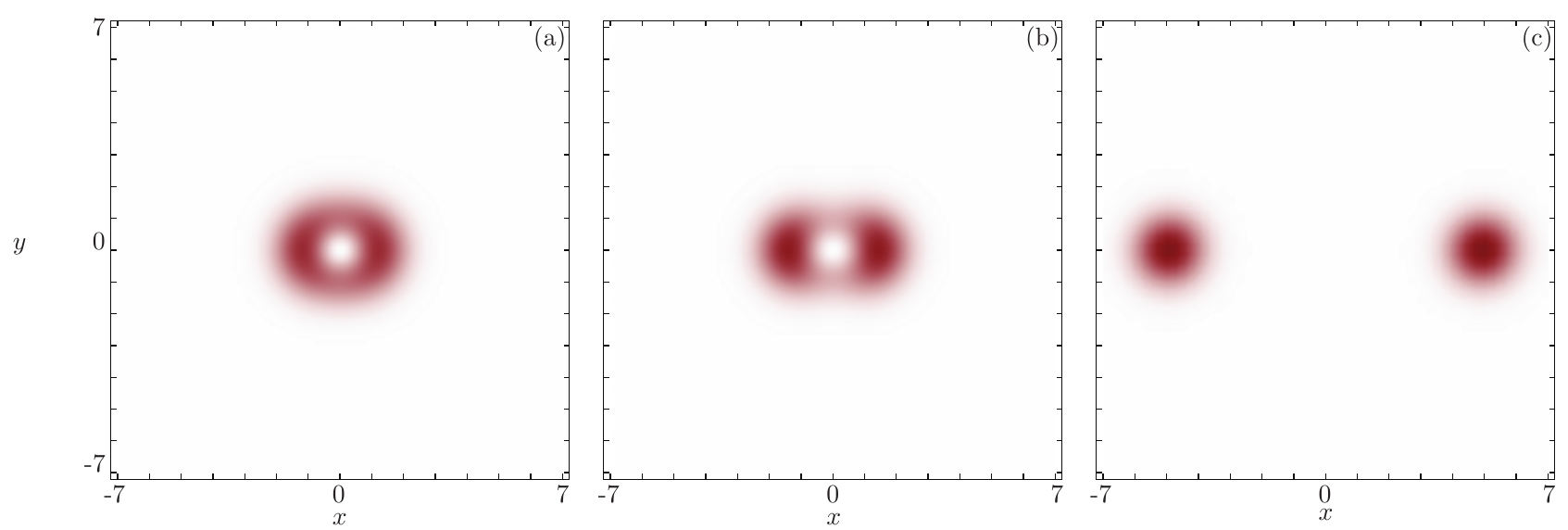}
	\caption{Husimi Q-function of the ground state for $\delta=$ (a) 0.01, (b) 0.5, and (c) 0.9;  all other parameters follow Fig.~\ref{fig:Fig1}.}\label{fig:Fig5}
\end{figure}

Finally, we follow the ground state of our model to see how the incoherent superposition of odd and even coherent states from the degenerate qubit regime, Fig. \ref{fig:Fig5}(a), starts displacing in the $x$-quadrature without noticeable changes in the $y$-quadrature, Fig. \ref{fig:Fig5}(b) and Fig. \ref{fig:Fig5}(c).

\section{conclusions}
We proposed a single parameter QRM that interpolates between the degenerate qubit and the relativistic regimes.
We provided the analytic solutions in these regimes.
The spectrum of the former is discrete and its eigenstates are odd and even displaced number states of the boson field correlated to the excited and ground state of the qubit, in that order.
The latter has continuum spectrum and its eigenstates are the unbalanced superposition of infinitely squeezed coherent states correlated to the qubit states.

We conducted a numerical statistical analysis below the transition from discrete to continuous spectrum regimes. 
We focused on the energy gap between first and second neighbors of the bosonic sector after Foulton-Guterman diagonalization in the qubit basis.
We found a structure that points to a distribution of first neighbors separation centered around two peaks with seemingly constant spectral separation for second neighbors.
Our statistical results seem to correlate with Braak conjecture regarding the pole and roots distribution for the full QRM.
We used Husimi Q-function to follow the process that displaces and squeezes the ground state from the degenerate qubit into the relativistic regime for reasonable control parameter values.

\begin{acknowledgments}
F.H.M.-V. acknowledges funding from CONACYT C\'atedra Grupal No. 551.
C.H.A. acknowledges funding from CONACYT doctoral grant No. 455378.
B.M.R.-L. acknowledges funding from CONACYT CB-2015-01-255230 grant and the Marcos Moshinsky Foundation Research Chair 2018.
\end{acknowledgments}

\providecommand{\noopsort}[1]{}\providecommand{\singleletter}[1]{#1}%
\end{document}